%% file: main.tex
\documentclass[conference]{IEEEtran} 
\usepackage{graphicx}
\usepackage{epsfig}
\usepackage{epstopdf}
\usepackage{amsmath,amsthm,amsfonts,amssymb}
\usepackage{multirow}
\usepackage{cite}
\usepackage{tikz}
\usepackage{ellipsis}
\usepackage{xcolor}
\usepackage{xifthen}
\usepackage{cancel}
\theoremstyle{definition} 
\theoremstyle{definition} \newtheorem{prop}{Proposition}
\theoremstyle{definition} 
\theoremstyle{definition}

\input{figures/commands.tex}
\begin{document}

\title{Optimizing MDS Codes for Caching at the Edge}

\author{\IEEEauthorblockN{Valerio Bioglio, Fr\'ed\'eric Gabry, Ingmar Land}
\IEEEauthorblockA{Mathematical and Algorithmic Sciences Lab\\ France Research Center, Huawei Technologies Co. Ltd.\\
Email: $\{$valerio.bioglio,frederic.gabry,ingmar.land$\}$@huawei.com}} 

\maketitle

\begin{abstract}
In this paper we investigate the problem of optimal MDS-encoded cache placement at the wireless edge to minimize the backhaul rate in heterogeneous networks. We derive the backhaul rate performance of any caching scheme based on file splitting and MDS encoding and we formulate the optimal caching scheme as a convex optimization problem. We then thoroughly investigate the performance of this optimal scheme for an important heterogeneous network scenario. We compare it to several other caching strategies and we analyze the influence of the system parameters, such as the popularity and size of the library files and the capabilities of the small-cell base stations, on the overall performance of our optimal caching strategy. Our results show that the careful placement of MDS-encoded content in caches at the wireless edge leads to a significant decrease of the load of the network backhaul and hence to a considerable performance enhancement of the network.

\end{abstract}

\section{Introduction}
\label{sec:intro}
\input{introduction.tex}

\section{System Model}
\label{sec:model}
\input{system-model.tex}

\section{Performance Analysis of Caching Schemes}
\label{sec:perf}
\input{results.tex}

\section{Numerical Illustrations}
\label{sec:num}
\input{numerical_illustrations.tex}

\section{Conclusions}
\label{sec:conclusions}
\input{future-work.tex}

\bibliographystyle{IEEEbib}
\bibliography{distributed_caching}

\end{document}

%% file: figures/commands.tex
\pgfmathsetseed{1}
\usetikzlibrary{shapes,positioning,arrows,calc,graphs}
\usetikzlibrary{decorations.pathreplacing,decorations.markings,shapes.geometric}
\tikzset{naming/.style={align=center,font=\small}}
\tikzset{antenna/.style={insert path={-- coordinate (ant#1) ++(0,0.25) -- +(135:0.25) + (0,0) -- +(45:0.25)}}}
\tikzset{station/.style={naming,draw,shape=dart,shape border rotate=90, minimum width=10mm, minimum height=10mm,outer sep=0pt,inner sep=3pt}}
\tikzset{mobile/.style={naming,draw,shape=rectangle,minimum width=12mm,minimum height=6mm, outer sep=0pt,inner sep=3pt}}
\tikzset{radiation/.style={{decorate,decoration={expanding waves,angle=90,segment length=4pt}}}}

\newcommand{\MBS}[1]{%
\begin{tikzpicture}
\node[station] (base) {#1};

\draw[line join=bevel] (base.100) -- (base.80) -- (base.110) -- (base.70) -- (base.north west) -- (base.north east);
\draw[line join=bevel] (base.100) -- (base.70) (base.110) -- (base.north east);

\draw[line cap=rect] ([xshift=-.1768cm,yshift=.6pt]base.north -| base.right tail) [antenna=1];
\draw[line cap=rect] ([yshift=.6pt]ant1 |- base.north) -- node[above,shape=rectangle,inner ysep=+.3333em]{\dots} ([xshift=.1768cm,yshift=.6pt]base.north -| base.left tail) [antenna=2];

\end{tikzpicture}
}

\newcommand{\BS}[1]{%
\begin{tikzpicture}
\node[station] (base) {#1};

\draw[line join=bevel] (base.100) -- (base.80) -- (base.110) -- (base.70) -- (base.north west) -- (base.north east);
\draw[line join=bevel] (base.100) -- (base.70) (base.110) -- (base.north east);

\draw[line cap=rect] ([yshift=0pt]base.north) [antenna=1];
\end{tikzpicture}
}

%% file: introduction.tex
Caching content at the wireless edge is a promising technique for future 5G wireless networks \cite{cache_magazine}.
One of the main motivations behind the idea of edge caching comes from the possibility of significantly reducing the backhaul usage and thus the latency in content retrieval by bringing the content closer to the end users. Exploiting the new capabilities of future multi-tier networks \cite{hetnets}, numerous works have recently investigated the potential benefits of caching data in densely deployed small-cell base stations (SBS) equipped with storage capabilities \cite{proactive}. 
In \cite{EEICC} it was shown that caching at the edge can provide significant gains in terms of energy efficiency which is considered a fundamental metric for future wireless networks.

 The investigation of caching from an information-theoretic point of view is presented in \cite{caching_networks}, where caching metrics are defined and analyzed for large networks. 
In \cite{modeling_tradeoffs} the problem is studied in terms of outage probability. 
Moreover, other embodiments and advantages of edge caching have been discussed in the literature. 
In \cite{exploit_mobility}, the idea of using the mobility of users in the network to increase caching gains is proposed, while in \cite{distributed_D2D} the authors leverage the possibility of exploiting the storage capabilities of mobile users by caching content directly on the users' devices.

In order to improve the theoretical performance limits of caching, a new interesting perspective stems from using ideas from coding theory. 
In \cite{fund_lim}, network coding (NC) is exploited in a scenario where each user is equipped with a cache. For this scenario, the authors show that a network coding based caching scheme decreases the backhaul load in comparison to the usual uncoded schemes. 
However, the extension of NC to a more complex scenario, where each user can be served by multiple SBSs, has been proved to be NP-complete \cite{exploit_NC}. 
Moreover, the use of NC in a scenario where the information is sent via unicast transmission does not give any advantage. 
Another promising related direction is the use of maximum-distance separable (MDS) codes to virtually increase the storage size in case of overlapping coverage areas. 
In \cite{femtocaching}, content is conveyed to distributed helpers using MDS codes to remedy the limitations of uncoded delivery in terms of delay. Assuming the network topology is known, the content placement strategy is formulated as a convex minimization of the overall delay. 
A random caching scheme for a cooperative multiple-input multiple-output (MIMO) network is combined with MDS encoding of the cached content to increase the gains resulting from MIMO cooperation \cite{opp_cache}. \\
Departing from those previous works, in this paper we investigate the performance of MDS-encoded distributed caching at the SBSs. In particular, we consider a scenario where mobile users can be served by multiple SBSs. 
In this setup, we derive the joint optimal encoding and placement of the encoded packets at the SBSs in order to minimize the backhaul rate.
Our main contributions are:

\begin{itemize}
\item We formally define the problem of caching MDS-encoded content at the wireless edge for a heterogeneous network (HetNet) scenario.
\item We derive the backhaul rate performance of a caching scheme based on storing MDS encoded packets, formulating the optimal caching scheme as a convex optimization problem.
\item We investigate the performance of the proposed packet placement and delivery scheme for a relevant HetNet scenario. 
Moreover, we compare it to several caching schemes, analyzing the benefits of MDS encoding on the performance of the scheme, as well as the importance of carefully optimizing the placement of the coded packets.
\item Our results show that the use of  MDS codes in the placement of content at the wireless edge yields to significant offloading of the network backhaul. 
\end{itemize}
This paper is organized as follows. 
In Section \ref{sec:model}, we define our system model, caching scheme and performance measures. 
In Section \ref{sec:perf}, we derive the main theoretical results of the paper. 
In Section \ref{sec:num} we thoroughly investigate the performance of our optimal schemes and we compare it to other caching schemes in a heterogeneous network scenario. 
Finally, Section \ref{sec:conclusions} concludes this paper.

\input{figures/geo_topology.tex}

%% file: figures/geo_topology.tex
\begin{figure}[ht!]
 \centering
\resizebox{0.45\textwidth}{!}{  \begin{tikzpicture}
    \coordinate (Origin)   at (0,0);
    \coordinate (XAxisMin) at (-5,0);
    \coordinate (XAxisMax) at (5,0);
    \coordinate (YAxisMin) at (0,-5);
    \coordinate (YAxisMax) at (0,5);
   \draw [thick,black] (0,0) circle (6cm);
 
\foreach \x in {1,...,10}{
\foreach \y in {1,...,10}{
\pgfmathparse{rand} \pgfmathsetmacro\xa{\pgfmathresult}
\pgfmathparse{rand}\pgfmathsetmacro\ya{\pgfmathresult} 
\draw [thick,fill=yellow!30] (5*\xa,5*\ya) circle (0.1cm);
}}

\node[red,thick,inner sep=0pt] (MbS){\MBS{MBS}} (0,0);
    \foreach \x in {-2,...,2}{
      \foreach \y in {-2,...,2}{
       \ifthenelse{\NOT 0 = \x \OR \NOT 0 = \y}{

        \node[scale=0.6] at (2*\x,2*\y) {\BS{SBS}} {};
        \draw[fill=none,dashed,green] (2*\x,2*\y) circle (1.7cm);
  \node  at (2*\x,2*\y) {};}{}
      }
    }   
    \filldraw[fill=gray, fill opacity=0.3, draw=black] (2,2)
        rectangle (4,0);
         \node[scale=0.6] at (0,6) {\BS{SBS}} {};
        \draw[fill=none,dashed,green] (0,6) circle (1.7cm);
  \node  at (0,6) {};
  
    \node[scale=0.6] at (0,-6) {\BS{SBS}} {};
        \draw[fill=none,dashed,green] (0,-6) circle (1.7cm);
  \node  at (0,-6) {};
  
    \node[scale=0.6] at (6,0) {\BS{SBS}} {};
        \draw[fill=none,dashed,green] (6,0) circle (1.7cm);
  \node  at (6,0) {};
  
    \node[scale=0.6] at (-6,0) {\BS{SBS}} {};
        \draw[fill=none,dashed,green] (-6,0) circle (1.7cm);
  \node  at (-6,0) {};
 \end{tikzpicture}}
  \caption{Heterogeneous network.}
  \label{fig:geo_topology}
\end{figure}
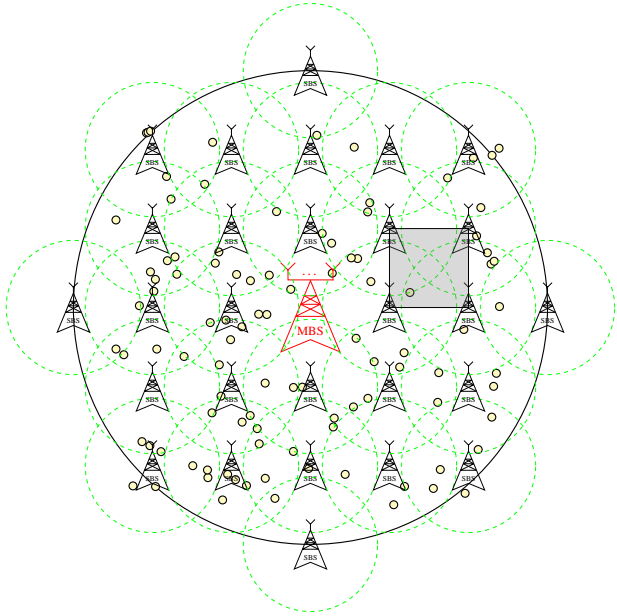

%% file: system-model.tex
In this section we describe our system model and we formally define caching schemes for heterogeneous networks.
\subsection{Network Model}
\label{sub:Net}
We consider the network illustrated in Fig.~\ref{fig:geo_topology}, where $U$ wireless users send requests to download files from a macro-cell base station.
The macro-cell base station (MBS) has access to a library of $N$ files $\mathcal{F} = \{ F_1, \dots, F_N \}$, each of size $B$ bits. 
Note that the assumption of equal size files is justifiable in practice since files can be divided into blocks of the same size.
We suppose that the users request entire files stored in the library, i.e., if $k_u$ is the data request for user $u$, then $k_u \in \mathcal{F}$. 
Each file has a different probability to be requested by the users. 
In particular, the probability distribution vector of the files is denoted by $p = \{ p_1, \dots, p_N \}$, referred to as the file popularity, where file $F_j$ is requested with probability $p_j$ and $\sum_{j=1}^{N} p_j = 1$.

Moreover  $N_{\text{SBS}}$ small-cell base stations are deployed in order to serve user requests within short distance. 
We assume that each SBS has a cache of size $M \cdot B$ bits (i.e., it can store up to $M < N$ complete files), and can send data to the users through an error-free link, i.e., we assume that content is delivered without errors as long as a user is within the coverage range. 
We denote by $r$ the coverage range of the $\text{SBS}$s. 
Each user requesting for files in $\mathcal{F}$ is initially served by the $\text{SBS}$s it can contact. 
If the requested file is not completely present in the caches, the $\text{MBS}$ has to send the missing data to the user, using the backhaul connection. 
The topology of the overall network may vary during time; in particular at each instant $t$, the connection network can be described as illustrated in Fig.~\ref{fig:graph}.

Each user $u \in \{ 1, \dots, U \}$ can be served by $d_u$ SBSs depending on its location in the area (see Fig.~\ref{fig:square}). 
In particular, we call $\gamma_i$ the probability for a user to be served by $d_u = i$ SBSs. 
This probability depends on the SBSs deployment and on the density of the users: in Section \ref{sec:num} we will show how to evaluate this probability in any scenario. 
\input{figures/graph.tex}

\subsection{Caching Schemes}
\label{sub:scheme}
A caching scheme is constituted of two phases, namely the \emph{placement phase} and the \emph{delivery phase}.
\begin{enumerate}
\item \emph{Placement phase}: In this phase, the caches of the $\text{SBS}$s are filled according to the placement strategy. 
This phase typically occurs at a moment with a low amount of file requests, e.g. at night.
\item \emph{Delivery phase}: In this phase, the users send requests to the $\text{MBS}$. 
These requests are initially served by the $\text{SBS}$s covering their locations. 
If a user cannot collect enough information to recover the file, the $\text{MBS}$ has to deliver the missing information through the backhaul. 
\end{enumerate}
Before the placement phase, each file in the library is split into $n$ \emph{fragments}, i.e., $F_j = \{ f_{1}^{(j)}, \cdots, f_{n}^{(j)} \}$ for all $1 \leq j \leq N$. 
In a coded caching scheme, these fragments are used to create $E_j$ \emph{encoded packets} $\{ e_{1}^{(j)}, \cdots, e_{E_j}^{(j)} \}$. 
We want to highlight that a fragment is an uncoded part of a file, while a packet is an encoded part of a file. 
In the following, a part of a file can denote either a fragment or an encoded packet.

In the placement phase, each $\text{SBS}$ receives $m_j$ parts of file $F_j$, where $\mathbf{m}=[m_{1} \cdots m_{N}]$ is referred to as the \emph{cache placement}. 
The \emph{placement problem} consists of finding the optimal number of parts $m_j$ of each file to be stored in the caches in order to minimize the average backhaul rate, which is defined as the average fraction of files that needs to be downloaded from the MBS during the delivery phase. 
\input{figures/square.tex}

%% file: figures/graph.tex
\begin{figure}[!t]
  \centering
\resizebox{0.47\textwidth}{!}{ \begin{tikzpicture}
[stack/.style={rectangle split, rectangle split parts=4, draw, anchor=center},
  myarrow/.style={single arrow, draw=none}]
  [->,>=stealth',shorten >=1pt,auto,
                    semithick]
\node[red,thick,inner sep=0pt,scale=0.8] (MBS){\MBS{MBS}} (0,2);
\node[stack,left= 2.2cm of MBS] (db) {$F_1$\nodepart{two}$F_2$\nodepart{three}$\cdots$\nodepart{four}$F_N$} (0,0);
\node[above= 0.1cm of db] {File Library};
\path (db.east) edge[thick,line width=0.7mm] node[above] {Core Network} (MBS.west);
\node[scale=0.6]  at (-3,-3) (SBS1) {\BS{SBS$_1$}};
\node[scale=0.6]  at (-1,-3) (SBS2) {\BS{SBS$_2$}};
\node[scale=0.6]  at (1,-3) (SBS3) {\BS{SBS$_3$}};
\node[scale=0.6]  at (3,-3) (SBS4) {\BS{SBS$_4$}};
\node[stack,left= 0.1cm of SBS1,scale=0.6] (db1) {$1$\nodepart{two}$2$\nodepart{three}$\cdots$\nodepart{four}$M$};
\node[stack,left= 0.1cm of SBS2,scale=0.6] (db2) {$1$\nodepart{two}$2$\nodepart{three}$\cdots$\nodepart{four}$M$};
\node[stack,right= 0.1cm of SBS3,scale=0.6] (db3) {$1$\nodepart{two}$2$\nodepart{three}$\cdots$\nodepart{four}$M$};
\node[stack,right= 0.1cm of SBS4,scale=0.6] (db4) {$1$\nodepart{two}$2$\nodepart{three}$\cdots$\nodepart{four}$M$};
\node[fill=yellow!20,draw=none,text=red,circle]  at (-4,-6) (u1) {u$_1$};
\node[fill=yellow!20,draw=none,text=orange,circle]  at (-3,-6) (u2) {u$_2$};
\node[fill=yellow!20,draw=none,text=blue,circle]  at (-2,-6) (u3) {u$_3$};
\node[fill=yellow!20,draw=none,text=orange,circle]  at (-1,-6) (u4) {u$_4$};
\node [fill=yellow!20,draw=none,text=orange,circle]  at (0,-6) (u5) {u$_5$};
\node[fill=yellow!20,draw=none,text=green,circle]  at (1,-6) (u6) {u$_6$};
\node[fill=yellow!20,draw=none,text=green,circle]  at (2,-6) (u7) {u$_7$};
\node[fill=yellow!20,draw=none,text=green,circle]  at (3,-6) (u8) {u$_8$};
\node[fill=yellow!20,draw=none,text=green,circle]  at (4,-6) (u9) {u$_9$};
\path (SBS1.north) edge[thick,line width=0.5mm] node[left=1cm] {$\cdots$} (MBS.south);
\path (SBS2.north) edge[thick,line width=0.5mm] node {} (MBS.south);
\path (SBS3.north) edge[thick,line width=0.5mm] node {} (MBS.south);
\path (SBS4.north) edge[thick,line width=0.5mm] node[right=1cm] {$\cdots$} (MBS.south);
\path (SBS1) edge[->,red] node[left=0.2cm,text=black] {$\cdots$} (u1);
\path (SBS1) edge[->,yellow!80] node {} (u2);
\path (SBS1) edge[->,yellow!80] node {} (u5);
\path (SBS1) edge[->,blue] node {} (u3);
\path (SBS2) edge[->,yellow!80] node {} (u2);
\path (SBS2) edge[->,blue] node {} (u3);
\path (SBS2) edge[->,yellow!80] node {} (u4);
\path (SBS2) edge[->,yellow!80] node {} (u5);
\path (SBS2) edge[->,green] node {} (u6);
\path (SBS3) edge[->,yellow!80] node {} (u2);
\path (SBS3) edge[->,yellow!80] node {} (u4);
\path (SBS3) edge[->,blue] node {} (u3);
\path (SBS3) edge[->,green] node {} (u7);
\path (SBS3) edge[->,green] node {} (u8);
\path (SBS3) edge[->,green] node {} (u9);
\path (SBS4) edge[->,blue] node {} (u3);
\path (SBS4) edge[->,yellow!80] node {} (u4);
\path (SBS4) edge[->,yellow!80] node {} (u5);
\path (SBS4) edge[->,green] node {} (u6);
\path (SBS4) edge[->,green] node {} (u7);
\path (SBS4) edge[->,green] node {} (u8);
\path (SBS4) edge[->,green] node[right=0.2cm,text=black] {$\cdots$} (u9);
\end{tikzpicture}}
 \caption{Instantaneous Heterogeneous network topology.}
  \label{fig:graph}
\end{figure}
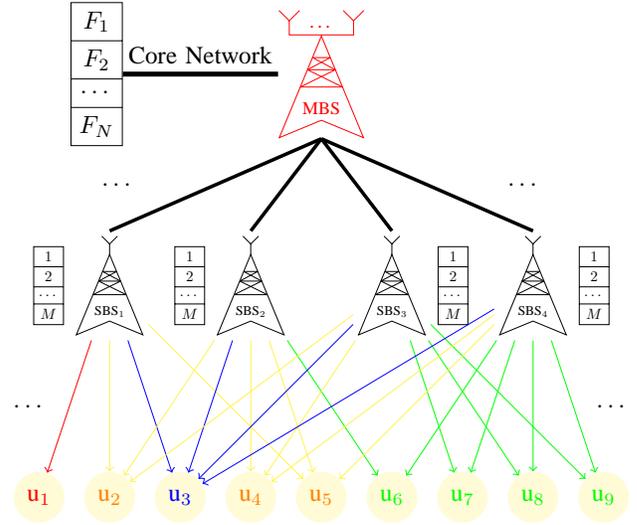

%% file: figures/square.tex
\def\firstcircle{(1,1) circle (1.7cm)}
\def\secondcircle{(1,-1) circle (1.7cm)}
\def\thirdcircle{(-1,-1) circle (1.7cm)}
\def\fourthcircle{(-1,1) circle (1.7cm)}
\def\rectangle{(-1,1) rectangle (1,-1)}
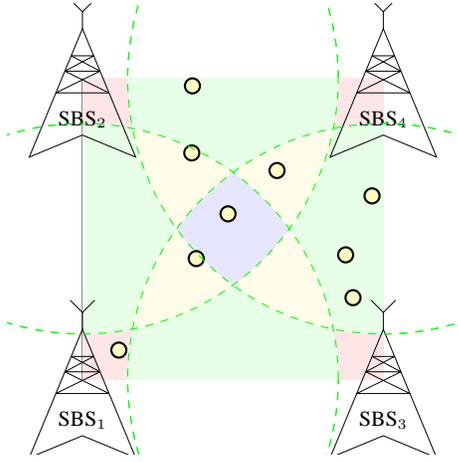
\begin{figure}[ht!]
  \centering
  \begin{tikzpicture}[scale=2]
  \clip (-1.5,-1.5) rectangle (1.5,1.5);
   \draw[black,fill=blue!20] (-1,1)
        rectangle (1,-1);

    \draw[fill=none,dashed,green] \firstcircle ;
    \draw[fill=none,dashed,green] \secondcircle;
    \draw[fill=none,dashed,green] \thirdcircle ;
 \draw[fill=none,dashed,green] \fourthcircle ;

    \begin{scope}
      \clip \rectangle;
      \fill[red!10] \firstcircle;
    \end{scope}
     \begin{scope}
      \clip \rectangle;
      \fill[red!10] \secondcircle;
    \end{scope}
     \begin{scope}
      \clip \rectangle;
      \fill[red!10] \thirdcircle;
    \end{scope}
    \begin{scope}
      \clip \rectangle;
      \fill[red!10] \fourthcircle;
    \end{scope}

    \begin{scope}
      \clip \rectangle;
      \clip \firstcircle;
      \fill[green!10] \secondcircle;
    \end{scope}

 \begin{scope}
      \clip \rectangle;
      \clip \firstcircle;
      \fill[green!10] \fourthcircle;
    \end{scope}
    
     \begin{scope}
      \clip \rectangle;
      \clip \thirdcircle;
      \fill[green!10] \secondcircle;
    \end{scope}
    
     \begin{scope}
      \clip \rectangle;
      \clip \thirdcircle;
      \fill[green!10] \fourthcircle;
    \end{scope}
    
     \begin{scope}
      \clip \rectangle;
      \clip \secondcircle;
      \clip \thirdcircle;
      \fill[yellow!10] \fourthcircle;
    \end{scope}
    
      \begin{scope}
      \clip \rectangle;
      \clip \secondcircle;
      \clip \thirdcircle;
      \fill[yellow!10] \firstcircle;
    \end{scope}
    
      \begin{scope}
      \clip \rectangle;
      \clip \firstcircle;
      \clip \fourthcircle;
      \fill[yellow!10] \secondcircle;
    \end{scope}
    
        \begin{scope}
      \clip \rectangle;
      \clip \firstcircle;
      \clip \fourthcircle;
      \fill[yellow!10] \thirdcircle;
    \end{scope}
    
          \begin{scope}
      \clip \rectangle;
      \clip \firstcircle;
      \clip \secondcircle;
      \clip \fourthcircle;
      \fill[blue!10] \thirdcircle;
    \end{scope}

\foreach \x in {1,...,3}{
\foreach \y in {1,...,3}{
\pgfmathparse{rand} \pgfmathsetmacro\xa{\pgfmathresult}
\pgfmathparse{rand}\pgfmathsetmacro\ya{\pgfmathresult} 
\draw [thick,fill=yellow!30] (\xa,\ya) circle (0.05cm);
}}
 
   \node[scale=0.9] at (-1,-1) {\BS{SBS$_1$}} {};
        \node[scale=0.9] at (-1,1) {\BS{SBS$_2$}} {};
        \node[scale=0.9] at (1,-1) {\BS{SBS$_3$}} {};
        \node[scale=0.9] at (1,1) {\BS{SBS$_4$}} {};
         \draw[fill=none,dashed,green] \firstcircle ;
    \draw[fill=none,dashed,green] \secondcircle;
    \draw[fill=none,dashed,green] \thirdcircle ;
 \draw[fill=none,dashed,green] \fourthcircle ;
   \end{tikzpicture}
  \caption{Small cells topology.}
  \label{fig:square}
\end{figure}

%% file: results.tex
In this paper, we propose to store MDS coded packets at the $\text{SBS}$s instead of fragments or entire files. 
We initially present in detail our scheme, and calculate the average backhaul rate of a MDS coded caching scheme. 
Afterwards, we prove that, given a placement scheme $\mathbf{m}=[m_{1} \cdots m_{N}]$, storing encoded packets always provides an advantage over storing fragments in terms of minimization of the backhaul rate. 
Finally, we formulate the MDS coded caching as an optimization problem, and show that finding an optimal solution for the problem is tractable. 


\subsection{MDS Coded Caching Scheme}
Maximum-distance separable codes are codes matching the singleton bound. 
In practice, using a MDS($a,b$) code it is possible to create $a$ encoded packets from $b$ input fragments, such that any subset of $b$ encoded packets is sufficient to recover the data. 
The most known example of MDS codes are the Reed-Solomon (RS) codes \cite{rs_book}.
Lately, sub-optimal MDS codes were proposed, like fountain codes \cite{fount_mckay}, where $b(1 + \epsilon)$ encoded packets are needed for the recovery.

In the proposed scheme, each $\text{SBS}$ receives $m_j$ different MDS coded packets for each file $F_j$ to be stored in its cache, with $\mathbf{m}=[m_{1} \cdots m_{N}]$. 
Moreover, the $\text{MBS}$ stores $n - m_j$ encoded packets for each file $F_j$ in order to serve the requests. 
Formally, we propose the following: 
\begin{enumerate}
\item \emph{Placement phase}: The $\text{MBS}$ creates $E_j = n + (N_{SBS} - 1)m_j$ encoded packets using a MDS($E_j,n$) code. 
The $\text{MBS}$ keeps $n - m_j$ encoded packets, and sends the other ones so that each $\text{SBS}$ stores $m_j$ unique packets. 
\item \emph{Delivery phase}: A user requesting file $F_j$ contacts $d_u \geq 1$ $\text{SBS}$s, receiving $m_j d_u$ different encoded packets. 
If $m_j d_u \geq n$, the user can recover the file due to the MDS-property of the code. Otherwise the $\text{MBS}$ sends the remaining $n - m_j d_u \leq n - m_j$ encoded packets. 
Since the $\text{MBS}$ kept $n - m_j$ encoded packets, the user does not receive replicated packets, and can decode the file. 
\end{enumerate}
In the following, we study the performance of the proposed MDS coded scheme in terms of average use of the backhaul link, and we compare the results with the uncoded scheme. 

\subsection{Performance Analysis}
To begin with, we calculate the average backhaul rate of the proposed MDS coded caching scheme. 
\begin{prop}
\label{prop:rate}
The average backhaul rate for an encoded caching placement scheme $\mathcal{C}_{\mathbf{m}}^{\text{MDS}}$ defined by the placement $\mathbf{m}=[m_{1} \cdots m_{N}]$ can be calculated as 
\begin{equation}
\label{eq:th_rate}
R_{(\mathcal{C}_{\mathbf{m}}^{\text{MDS}})} = \sum_{d=1}^{S} \sum_{j=1}^{N} \gamma_i p_j \left( 1 - \min \left( 1, \frac{d m_j}{n} \right) \right),
\end{equation}
where $S \leq N_{\text{SBS}}$ is the maximum number of $\text{SBS}$s serving a user
\end{prop}
\begin{IEEEproof}
Let $u$ be a user served by $d_u$ $\text{SBS}$s.
If $u$ requests file $F_j$, exactly $n$ packets have to be collected in order to retrieve $F_j$. 
Each $\text{SBS}$ stores $m_j$ packets of the file $F_j$, hence the user can collect $d_u  m_j$ different encoded packets. 
If $d_u  m_j \geq n$ the $\text{MBS}$ does not send any packet, otherwise $n - d_u  m_j$ packets are sent through the backhaul. 
The rate for user $u$ requesting for file $F_j$ can consequently be calculated as 
\begin{equation}
\label{eq:init}
R(F_j)= 1 - \min \left( 1, \frac{d_u m_j}{n} \right).
\end{equation}
A user has the probability $\gamma_d$ to be served by $d$ $\text{SBS}$s, hence the expected rate for the file $F_j$ is given by
\begin{equation*}
\bar{R}(F_j) = \sum_{d=1}^{S} \gamma_d \left( 1 - \min \left( 1, \frac{d m_j}{n} \right) \right).
\end{equation*}
Finally, each file has a different probability to be requested by a user. 
Averaging over the request probability distribution $p = [p_{1} \cdots p_{N}]$, The average backhaul rate for the encoded caching placement scheme is given by
\begin{equation*}
R_{(\mathcal{C}_{\mathbf{m}}^{\text{MDS}})} = \sum_{d=1}^{S} \sum_{j=1}^{N} \gamma_d p_j \left( 1 - \min \left( 1, \frac{d m_j}{n} \right) \right).
\end{equation*}
\end{IEEEproof}
In order to evaluate the gain given by the use of MDS codes, we calculate the backhaul rate of a random caching scheme, where only file fragmentation is exploited. 
In practice, given a placement $\mathbf{m} = [m_{1} \cdots m_{N}]$, the $\text{MBS}$ sends to each $\text{SBS}$ $m_j$ different fragments randomly drawn among the $n$ fragments of file $F_j$. 
\begin{prop}
\label{prop:rate_uncoded}
The average backhaul rate for a random caching placement scheme $\mathcal{C}_{\mathbf{m}}$ defined by the placement $\mathbf{m}=[m_{1} \cdots m_{N}]$ can be calculated as 
\begin{equation}
\label{eq:th_rate_uncoded}
R_{(\mathcal{C}_{\mathbf{m}})} = \sum_{d=1}^{S} \sum_{j=1}^{N} \gamma_d p_j \left( 1 - \frac{m_j}{n} \right)^d.
\end{equation}
\end{prop}
\begin{IEEEproof}
The proof is similar to the one proposed for Proposition~\ref{prop:rate}. 
The main difference is the calculation of the rate for a user requesting for a file. 
For the given scheme, if a user $u$ requesting file $F_j$ is served by $d_u$ $\text{SBS}$s, the average backhaul rate is
\begin{equation}
\label{eq:init_uncoded}
\left( 1 - \frac{m_j}{n} \right)^{d_u}.
\end{equation}
This is due the fact that in this case the fragments are not unique since they can be replicated in different caches. Therefore the probability that a single fragment is not present in any of the caches is $(1 - m_j/n)^{d_u}$. The rest of the proof is similar to the one of Proposition~\ref{prop:rate}. 
\end{IEEEproof}
Now we can prove that, given a cache placement $\mathbf{m}=[m_{1} \cdots m_{N}]$, it is preferable in terms of backhaul usage to store MDS coded packets rather than fragments. 
\begin{prop}
\label{prop:comparison}
For any placement $\mathbf{m}=[m_{1} \cdots m_{N}]$, it holds that 
\begin{equation}
\label{eq:comp}
R_{(\mathcal{C}_{\mathbf{m}}^{\text{MDS}})} \leq R_{(\mathcal{C}_{\mathbf{m}})}.
\end{equation}
\end{prop}
\begin{IEEEproof}
As stated in the proof of Proposition~\ref{prop:rate_uncoded}, the difference between Equations (\ref{eq:th_rate}) and (\ref{eq:th_rate_uncoded}) is given by (\ref{eq:init}) and (\ref{eq:init_uncoded}) respectively. 
In order to prove the proposition, we have to prove that (\ref{eq:init_uncoded}) is larger than or equal to (\ref{eq:init}). \\ 
Since $d \geq 1$ and $0 \leq m_j/n \leq 1$, from Bernoulli's inequality
\begin{equation*}
\left( 1 - \frac{m_j}{n} \right)^{d} \geq 1 - d \frac{m_j}{n},
\end{equation*}
and obviously
\begin{equation*}
\left( 1 - \frac{m_j}{n} \right)^{d} \geq 0.
\end{equation*}
We conclude the proof by noticing that 
\begin{equation*}
 1 - \min \left( 1, \frac{d m_j}{n} \right) =  \min \left( 0, 1 - \frac{d m_j}{n} \right) \leq \left( 1 - \frac{m_j}{n} \right)^{d}.
\end{equation*}
\end{IEEEproof}

\subsection{Optimal MDS Coded Caching}
In the following, we study the placement problem of MDS encoded packets as an optimization problem, and show its tractability. 
Based on the average backhaul rate of Proposition \ref{prop:rate}, we have the following result.
\begin{prop}
Finding the optimal MDS coded placement scheme $\mathcal{C}_{(\text{opt})}^{\text{MDS}}$ defined by $\mathbf{m}_{(\text{opt})} = [m_{1} \cdots m_{N}]$, which minimizes the average backhaul rate $R_{(\mathcal{C}_{\mathbf{m}}^{\text{MDS}})}$, is a convex optimization problem.
\end{prop}
\begin{IEEEproof}
The placement problem can be recast as the following optimization problem:
\begin{equation}
\begin{array}{l}
\displaystyle 
\min_{m_1,\dots,m_N} \sum_{d=1}^{S} \sum_{j=1}^{N} \gamma_d p_j \left( 1 - \min \left( 1, \frac{d m_j}{n} \right) \right) \\
\displaystyle 
\text{s.t.} \: \sum_{j=1}^{N} m_j = M n \\
\:\: 0 \leq m_j \leq n \qquad \forall j = 1, \dots, N
\end{array}
\end{equation}
We recall that the number of fragments $n$ can be chosen, hence we can reformulate the optimization problem as
\begin{equation}
\begin{array}{l}
\displaystyle 
\min_{q_1,\dots,q_N} \sum_{d=1}^{S} \sum_{j=1}^{N} \gamma_d p_j ( 1 - \min ( 1, d q_j ) ) \\
\displaystyle 
\text{s.t.} \: \sum_{j=1}^{N} q_j = M \\
\:\: 0 \leq q_j \leq 1 \qquad \forall j \in [1,N]
\end{array}
\label{eq:conv}
\end{equation}
where $q_j = m_j/n$. 
It can be easily shown that \eqref{eq:conv} is a convex optimization problem, which concludes the proof.
\end{IEEEproof}

%% file: numerical_illustrations.tex
In this section we investigate the performance of the optimal MDS coded caching scheme in terms of backhaul rate in a HetNet topology of particular relevance. 
We emphasize that our numerical results can be further generalized to any network topology. 
In particular the challenging problem of the optimization of the locations of the small-cell base stations is of considerable practical and theoretical interest, but outside the scope of this paper. 

\subsection{Network Topology}
\label{num:topology}
In the following we consider the network depicted in Fig.~\ref{fig:geo_topology}, where the $\text{SBS}$s are deployed according to a regular grid with a distance $d=60$ meters between each $\text{SBS}$. 
The macro-cell base station coverage area $\mathcal{R}$ has a radius of $D=500$ meters. 
In order to reach any user in $\mathcal{R}$, each $\text{SBS}$ has a coverage area of radius $r$ such that  $ d/\sqrt{2} \leq r \leq d$, which means that the coverage areas are overlapping as shown in the highlighted square in Fig.~ \ref{fig:square}.

If we denote by $\mathcal{A}_d$ and $\rho_d$ the total area of $\mathcal{R}$ where a user can be served by $d$ $\text{SBS}$s and its average density, respectively, the probability $\gamma_d$ that a user is served by $d$ $\text{SBS}$s can be formally calculated as
\begin{equation*}
\gamma_d = \frac{\rho_d \mathcal{A}_d}{\sum_{i=1}^{S} \rho_i \mathcal{A}_i}.
\end{equation*}
We note that for this particular deployment, the coefficients $\mathcal{A}_d$ can be theoretically approximated using simple geometrical calculations.

In the following, we consider the users to be uniformly distributed in $\mathcal{R}$, with density $\rho_d = \rho=0.05$ users$/m^{2}$.
These numbers correspond to $316$ small-cell base stations being deployed, covering $U=31,415$ mobile users. 
The request probability of the files $p=[p_{1} \cdots p_{N}]$ is distributed according to a Zipf law of parameter $\alpha$, i.e.,
\begin{equation*}
p_{j}=\frac{1/j^{\alpha}}{\sum_j 1/j^{\alpha}},
\end{equation*}
where $\alpha$ represents the skewness of the distribution and can takes values in $[0.5;1.5]$, see e.g. \cite{zipf}. 

\subsection{Caching Strategies}
In order to evaluate the performance of the optimal MDS coded placement scheme, we compare the optimal achievable backhaul rate $R^{\text{MDS}}_{(\text{opt})}$  to three other practical placement schemes, with and without MDS encoding of the files: 
\begin{enumerate}
\item ``most popular'' placement $\mathcal{C}_{(\text{pop})}$ and $\mathcal{C}^{\text{MDS}}_{(\text{pop})}$: each $\text{SBS}$ stores the $M$ most popular files. 
Since entire files are stored for this scheme, we note that  $\mathcal{C}_{(\text{pop})}$ and $\mathcal{C}^{\text{MDS}}_{(\text{pop})}$ are equivalent. 
Hence we call $R^{\text{MDS}}_{(\text{pop})}$ the rate achieved using this scheme.
\item uniform placement $\mathcal{C}_{(\text{unif})}$ and $\mathcal{C}^{\text{MDS}}_{(\text{unif})}$: the file is divided into $n$ fragments, which are placed uniformly at random at each $\text{SBS}$, i.e., each $\text{SBS}$ stores $M/N$ fragments of each file. 
In the case that MDS codes are employed, the $\text{SBS}$s store $M/N$ encoded packets for each file instead of fragments. 
We call $R^{\text{MDS}}_{(\text{unif})}$ and $R_{(\text{unif})}$ the rates achieved using this scheme with and without MDS encoding, respectively.
\item proportional placement $\mathcal{C}_{(\text{prop})}$ and $\mathcal{C}^{\text{MDS}}_{(\text{prop})}$: for each file $F_j$, $n$ fragments are created, along with a congruous number of encoded packets. 
The amount of parts $m_j$ of file $F_j$ stored at the $\text{SBS}$s is proportional to its popularity $p_j$ while satisfying the size constraints. 
In particular, the number of parts cached at the $\text{SBS}$ is iteratively calculated as
\begin{equation*}
m_i = \min \left( n, p_i \cdot \frac{\sum_{j=1}^{i-1} m_j}{Mn} \right).
\end{equation*}
We call $R^{\text{MDS}}_{(\text{prop})}$  and $R_{(\text{prop})}$ the rates achieved using this scheme with and without MDS encoding, respectively.
\end{enumerate}

\begin{figure}[!t]
\centering
\includegraphics[scale=0.49]{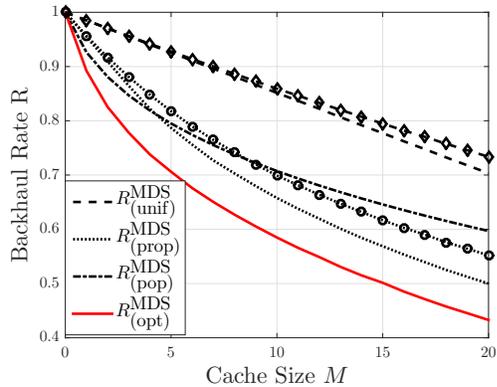}
\caption{Backhaul rate as a function of the cache size M, with N$=200$ files, $\alpha=0.7$ and $r=60$ meters. The lines without markers depict the use of MDS codes (encoded packets) in the placement, while the line with markers depict uncoded (fragments) placement.}
\label{fig:Cache}
\end{figure}

\subsection{Numerical Results}

In Fig.~\ref{fig:Cache} we depict the backhaul rates as a function of the SBS cache size $M$. 
To begin with, it can be noticed that the use of MDS codes gives an advantage in terms of reduction of the backhaul usage, since MDS placement always outperforms the uncoded placement for each of the schemes. 
This behavior confirms the result presented in Proposition~\ref{prop:comparison}. 
Consequently, in the remainder of our analysis we only show the rates of MDS caching schemes. 
Secondly, we observe that, as it can be expected, the backhaul rates are decreasing when the storage capacity of the $\text{SBS}$s increases and that the optimal caching scheme outperforms the three other schemes. 
Furthermore, we observe that the difference between $R_{(\text{opt})}^{\text{MDS}}$ and the rate of the other schemes increases as the cache size increases. 
Finally, we can notice that the (pop) scheme performs the closest to the optimal scheme for small values of $M$, getting worse as $M$ increases. 
We guess that this scheme is not able to really take advantage of the nature of future HetNets, i.e., of the overlapping between coverage areas of the $\text{SBS}$s. 

\begin{figure}[!t]
\centering
\includegraphics[scale=0.49]{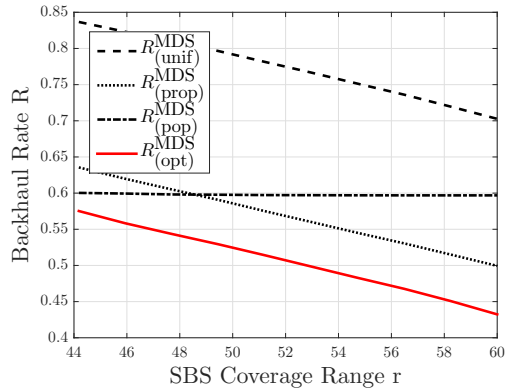}
\caption{Backhaul rate as a function of the SBS coverage radius r, with N$=200$ files, $\alpha=0.7$, and $M=20$ files.}
\label{fig:Range}
\end{figure}

In Fig.~\ref{fig:Range} we confirm the previous conjecture by representing the backhaul rates as a function of the coverage radius $r$ of the $\text{SBS}$s. 
We identify in the figure a striking behavior for the $\mathcal{C}_{(\text{pop})}^{\text{MDS}}$ placement scheme, as its rate does not depend on the coverage radius. 
This behavior is due to the fact that the most popular entire files are stored at the $\text{SBS}$s for this scheme: being served by more $\text{SBS}$s does not increase the probability of having access to new files since the same files are stored in every cache. 
In contrast to the $\mathcal{C}_{(\text{pop})}^{\text{MDS}}$ scheme, the performance of other schemes increases as the coverage area increases, and noticeably the optimal scheme significantly outperforms the other schemes as $r$ grows. 
As a numerical illustration of this fact, we can imagine the scenario where the $N = 200$ files are videos of size $100$Mbits. 
When $r=60$, the rate difference between $R_{\text{(opt)}}^{\text{MDS}}$ and the second best scheme $R_{\text{(prop)}}^{\text{MDS}}$ is $0.07$. 
This represents a difference of backhaul load of $7$Mbits$/s$ if there is $1$ demand per second in the macro cell, which is a considerable gain for current maximal backhaul capacities.

\begin{figure}[!t]
\centering
\includegraphics[scale=0.49]{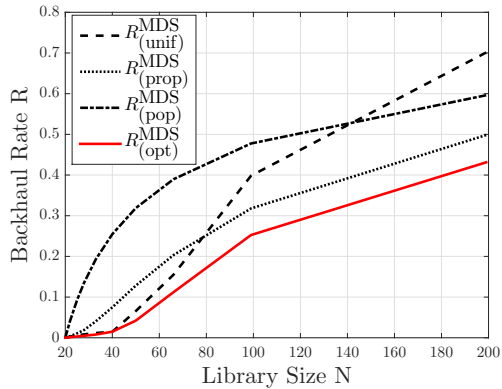}
\caption{Backhaul rate as a function of the library size N, with $\alpha=0.7$, M$=20$ files and $r=60$ meters.}
\label{fig:Size}
\end{figure}

Finally in Fig.~\ref{fig:Size} we illustrate the backhaul rates as a function of the library size $N$ while the size of the edge caches remains constant. 
For small library sizes, the $\mathcal{C}_{(\text{pop})}^{\text{MDS}}$ scheme is outperformed by the other schemes as a spreading of the library files over the caches is more efficient than only storing a few number of entire files. 
On the other hand, as the library size increases, the performance of the $\mathcal{C}_{(\text{pop})}^{\text{MDS}}$ scheme becomes better in comparison to the other schemes. 
The $\mathcal{C}_{(\text{pop})}^{\text{MDS}}$ scheme, however, is still outperformed by the $\mathcal{C}_{(\text{prop})}^{\text{MDS}}$ scheme, since we consider the overlapping scenario with coverage radius $r=60$ meters, which is unfavorable to the $\mathcal{C}_{(\text{pop})}^{\text{MDS}}$ scheme. 
Moreover, we can notice that initially the $\mathcal{C}_{(\text{unif})}^{\text{MDS}}$ scheme is really close to the optimal one. 
This is due to the fact that if the library size $N$ is small compared to the cache size $M$, the optimal scheme is well approximated by the scheme $\mathcal{C}_{(\text{unif})}^{\text{MDS}}$ that stores the packets uniformly across the files. 
As the library size increases, this na\"{i}ve scheme gets worse, since it does not take into account the popularity of the files. 
On the contrary, the performance of the $\mathcal{C}_{(\text{prop})}^{\text{MDS}}$ gets better, even if the gap with the optimal scheme becomes constant. 
This is due to the fact that this scheme does not truly exploit the possibility a user has to be served by more than one $\text{SBS}$. 
In general, the presence of a gap between the optimal scheme and sub-optimal ones highlights the importance of implementing the optimal caching placement scheme to exploit the overlapping coverage regions of the HetNet and thus to effectively decrease the backhaul load of the overall network. \\

%% file: future-work.tex
We considered the problem of optimal MDS-encoded content placement at the cache-equipped small-cell base stations at the wireless edge in order to minimize the overall backhaul load of the network. We derived the optimal caching placement strategy based on file splitting combined with MDS encoding, which we formulated as a convex optimization problem. We then deeply investigated the performance of this optimal placement for a relevant HetNet scenario by comparing it to existing caching strategies and by measuring the influence of the key parameters, such as the capabilities of the small-cell base stations and the library statistics. Our numerical observations, which can be easily generalized for any geometric topology, showed that optimizing the placement of MDS encoded packets at the wireless edge is of crucial importance, since it yields a significant decrease of the load of the network backhaul, and thus achieves a considerable improvement in terms of delay for content delivery to the end users.